\documentclass[sn-basic,Numbered]{sn-jnl}

\usepackage{graphicx}%
\usepackage{multirow}%
\usepackage{amsmath,amssymb,amsfonts}%
\usepackage{amsthm}%
\usepackage{mathrsfs}%
\usepackage[title]{appendix}%
\usepackage{xcolor}%
\usepackage{textcomp}%
\usepackage{manyfoot}%
\usepackage{booktabs}%
\usepackage{algorithm}%
\usepackage{algorithmicx}%
\usepackage{algpseudocode}%
\usepackage{listings}%

\begin{document}

\preprint{MS-TP-23-50}

\title[Quantum kinetic approach to the Schwinger production of scalar particles]{Quantum kinetic approach to the Schwinger production of scalar particles in an expanding universe}

\author*[1]{\fnm{Anastasia V.} \sur{Lysenko}}\email{anastasia.lysenko.fizfak.knu@gmail.com}

\author[1,2]{\fnm{Oleksandr O.} \sur{Sobol}}\email{oleksandr.sobol@knu.ua}

\affil[1]{\orgdiv{Physics Faculty}, \orgname{Taras Shevchenko National University of Kyiv}, \orgaddress{\street{64/13,~Volodymyrska Street}, \city{Kyiv}, \postcode{01601}, \country{Ukraine}}}

\affil[2]{\orgdiv{Institute for Theoretical Physics}, \orgname{University of M\"{u}nster}, \orgaddress{\street{Wilhelm-Klemm-Stra{\ss}e 9}, \city{M\"{u}nster}, \postcode{48149}, \country{Germany}}}


\abstract{We study the Schwinger pair creation of scalar charged particles by a homogeneous electric field in an expanding universe in the quantum kinetic approach. 
We introduce an adiabatic vacuum for the scalar field based on the Wentzel--Kramers--Brillouin solution to the mode equation in conformal time and apply the formalism of Bogolyubov coefficients to derive a system of quantum Vlasov equations for three real kinetic functions. Compared to the analogous system of equations previously reported in the literature, the new one has two advantages. First, its solutions exhibit a faster decrease at large momenta which makes it more suitable for numerical computations. Second, it predicts no particle creation in the case of conformally coupled massless scalar field in the vanishing electric field, i.e., it respects the conformal symmetry of the system.
We identify the ultraviolet divergences in the electric current and energy--momentum tensor of produced particles and introduce the corresponding counterterms in order to cancel them.}


\maketitle

\section{Introduction}
\label{sec:intro}

One of the most fascinating phenomena occurring in a strong electromagnetic field is the Schwinger effect or the creation of particle--antiparticle pairs from physical vacuum~\cite{Sauter:1931zz,Heisenberg:1936nmg,Schwinger:1951nm}. It was predicted almost a hundred years ago and until the present day has not been observed in the laboratory because of extremely high value of the required electric field $E\sim 10^{18}\,\text{V}/\text{m}$~\cite{Cohen:2008wz}. Nevertheless, such enormous fields might exist in the universe, e.g., around the compact objects like neutron stars or black holes~\cite{Ruffini:2009hg,Kim:2023qdj} or during the early stages of evolution~(see, e.g., Refs.~\cite{Kobayashi:2014zza,Sharma:2017eps,Domcke:2019qmm}).

The Schwinger effect in the early universe is often considered within the context of inflationary magnetogenesis models which include the coupling of the electromagnetic field to the inflaton or to the spacetime curvature~\cite{Turner:1987bw,Ratra:1991bn,Garretson:1992vt}. The simplest scenario for the Schwinger pair production during inflation which can be studied analytically is the case of a constant and homogeneous electric field (with possible presence of a collinear magnetic field) in de Sitter spacetime which has been studied in much detail in the literature~\cite{Kobayashi:2014zza,Frob:2014zka,Bavarsad:2016cxh,Stahl:2015gaa,Hayashinaka:2016dnt,Hayashinaka:2016qqn,Sharma:2017ivh,Bavarsad:2017oyv,Hayashinaka:2018amz,Banyeres:2018aax,Domcke:2018eki}. The results of this simple approach has been used for a lot of phenomenological applications~\cite{Tangarife:2017rgl,Sharma:2017eps,Stahl:2018idd,Geng:2017zad,Giovannini:2018qbq,Kitamoto:2018htg,Chua:2018dqh,Shakeri:2019mnt,Domcke:2019qmm,Sobol:2018djj,Sobol:2019xls,Gorbar:2021rlt,Gorbar:2021zlr}; however, in realistic inflationary models both underlying assumptions, the constant in time gauge field and purely de Sitter expansion of the universe, typically do not hold. 

In a time-dependent electromagnetic field, the Schwinger pair production has a more complicated dynamics. This was demonstrated by using the kinetic approach both in Minkowski space~\cite{Kluger:1991ib,Kluger:1992gb,Schmidt:1998vi,Kluger:1998bm,Schmidt:1998zh,Bloch:1999eu,Alkofer:2001ik,Kim:2011jw,Blaschke:2019pnj} and in the expanding universe~\cite{Gorbar:2019fpj,Sobol:2020frh}. In contrast to the case of constant field, the Schwinger induced current exhibits the non-Markovian character depending not only on the field at a given moment of time but also on the prehistory. Moreover, due to inertial properties of charge carriers the current is retarded with respect to the changes in the electric field which leads to the oscillatory behavior of the induced current and the electric field.

In the kinetic approach, the particle creation is described by the Schwinger source term which can be constructed phenomenologically~\cite{Kluger:1991ib,Kluger:1992gb,Gorbar:2019fpj} or derived from the first principles~\cite{Schmidt:1998vi,Kluger:1998bm,Schmidt:1998zh,Bloch:1999eu,Alkofer:2001ik,Kim:2011jw,Blaschke:2019pnj,Sobol:2020frh}. The latter way, although being much more involved, allows to capture the effects of particle statistics and the nonlocal in time character of the pair-creation process. There is, however, a fundamental issue which makes the application of such a first-principles approach ambiguous in the case of the expanding universe. This is a well-known in the literature problem of defining the vacuum state and particles in nonstationary backgrounds which do not match the Minkowski spacetime in the asymptotic past and future~\cite{Birrell-book,Parker-book,Parker:1968mv,Parker:1969au}. The result for the number of produced particles depends on the definition of physical vacuum state or, in other words, on the choice of the observer.

In this work, we revisit the problem of pair creation of the scalar particles by a time-dependent electric field in the expanding Friedmann--Lema\^{i}tre--Robertson--Walker (FLRW) universe previously considered in Ref.~\cite{Sobol:2020frh}. We mostly follow the same strategy in order to derive the system of quantum kinetic equations describing the Schwinger pair production in this system. However, in the present study, we perform all computations in terms of the conformal time rather then physical time considered in Ref.~\cite{Sobol:2020frh}. Despite the seemingly trivial change, we find that the definition of the adiabatic vacuum is different compared to the previous work and, consequently, the final system of equations appears to be different. We analyze the ultraviolet (UV) behavior of its solutions and show that kinetic functions exhibit faster decrease at large momenta as compared to the system in Ref.~\cite{Sobol:2020frh} which makes the new system of equations more attractive for numerical analysis. Also the new setup respects the conformal symmetry of the problem and does not lead to the particle production in the case of conformally coupled massless scalar field in the absence of the electric field.

The aim of this work is not to show that the computations in Ref.~\cite{Sobol:2020frh} are incorrect, but to demonstrate that there is a different and more convenient way to derive the system of quantum kinetic equations. Although we expect that the numerical result in the case of a strong electric field $|e\boldsymbol{E}|\gg H^2,\,m^2$ (where $e$ is the electric charge, $H$ is the Hubble parameter, and $m$ is the particle's mass) would be the same for both approaches, there might be some numerical differences in the case of a weak field where the particle production by the electric field and the time-varying metric are comparable.

The rest of the article has the following structure. In Sec.~\ref{sec:vlasov-eq} we consider the evolution of a quantum scalar field on a classical background of the electric field and the expanding universe. In particular, we define the adiabatic vacuum and use the formalism of the Bogolyubov coefficients in order to derive the system of quantum kinetic equations. Then, in Sec.~\ref{sec:observables} we determine the basic observables---electric current and energy--momentum tensor of the produced particles---and express them in terms of the kinetic functions. In Sec.~\ref{sec:renormalization}, we realize the renormalization program in order to separate the UV divergent contributions to the observables and cancel them by introducing the counterterms to the action. Section~\ref{sec:conclusions} is devoted to conclusions. In Appendix~\ref{app-expansion}, we determine the asymptotical behavior of the kinetic functions in the limit of large momenta. 

Throughout the work, we use natural units and set $\hbar=c=1$. We assume that the universe is described by a spatially flat FLRW metric. In terms of conformal time $\eta$ it has the form $g_{\mu\nu}=\operatorname{diag}(a^2,\,-a^2,\,-a^2,\,-a^2)$, where $a=a(\eta)$ is the scale factor.

\section{Quantum Vlasov equations for the scalar field}
\label{sec:vlasov-eq}

In order to describe the process of the Schwinger production of scalar charged particles by a strong electric field in the expanding universe we consider the action
\begin{equation}
\label{S}
S=\int d^{4}x\sqrt{-g} \left[-\frac{M_{p}^{2}}{2}R + \mathcal{L}_{{\rm EM}} + \mathcal{L}_{{\rm ch}} \right], 
\end{equation}
where  $g={\rm det}(g_{\mu\nu})=-a^8$ is the determinant of the spacetime metric, $R=-6a''/a^3$ is the Ricci curvature scalar for FLRW metric. Thus, the first term in the action represents the Einstein--Hilbert action for gravity. The second term in brackets in Eq.~\eqref{S} represents the Lagrangian density for the electromagnetic field\footnote{Throughout this work, by ``electromagnetic field'' we mean any Abelian gauge field, not necessarily the one corresponding to $U(1)_{\mathrm{EM}}$ subgroup of the Standard Model.}
\begin{equation}
\label{L_EM}
\mathcal{L}_{\mathrm{EM}}=-\frac{1}{4}F_{\mu\nu}F^{\mu\nu}+\mathcal{L}_{\mathrm{int}}(A_{\mu},\phi),
\end{equation}
where the term $\mathcal{L}_{\mathrm{int}}$ describes the coupling of the electromagnetic field to some generic field $\phi$. Although this interaction is necessary to produce sufficiently strong electromagnetic field needed for the Schwinger pair production, in this work we will not stick to any specific model and will assume that the electromagnetic field already exists in the universe. Moreover, we consider the following configuration of the gauge field: (i) the magnetic field $\boldsymbol{B}$ is negligibly small compared to the electric one $\boldsymbol{E}$; (ii) the coherence length of the electric field $\lambda_E$ is much greater than any other physically relevant length scale in the problem, e.g., the Schwinger pair production scale $l_{\mathrm{S}}\sim |e\boldsymbol{E}|^{-1/2}$ or the Hubble scale $l_H\sim H^{-1}$. The latter condition allows us to treat the electric field spatially homogeneous and dependent only on time $\boldsymbol{E}=\boldsymbol{E}(\eta)$. Such a field configuration can be described by the vector potential (in the Coulomb gauge) $A_{\mu}=(0,\,\boldsymbol{A}(\eta))$ with $\boldsymbol{A}'(\eta)=-a^2 \boldsymbol{E}$. With the factor $a^2$, the electric field $\boldsymbol{E}$ is a physical field measured by the comoving observer.

Finally, we go back to the last term in Eq.~\eqref{S} which represents the Lagrangian density of the complex scalar charged field $\chi$ with mass $m$ and charge $e$:
\begin{equation}
\label{L_ch}
\mathcal{L}_{{\rm ch}}=g^{\mu\nu}(\mathcal{D}_{\mu}\chi)^{\dagger}(\mathcal{D}_{\nu}\chi)-\left( m^{2}-\xi R \right) \lvert\chi\rvert^{2}.
\end{equation}
Here $\lvert\chi\rvert^{2}\equiv \chi^{\dagger}\chi$, $\mathcal{D}_{\mu}=\partial_{\mu}-ieA_{\mu}$ is the covariant derivative acting on the scalar field, and $\xi$ is the coupling constant responsible for the nonminimal coupling of the scalar field $\chi$ to gravity.

Varying action in Eq.~\eqref{S} with respect to $\chi^{\dagger}$, we obtain the equation of motion for the scalar field $\chi$:
\begin{equation}
\label{Lag_Eyl}
\frac{1}{\sqrt{-g}}\mathcal{D}_{\mu}\left[\sqrt{-g}g^{\mu\nu}\mathcal{D}_{\nu}\chi \right] + \left( m^{2}-\xi R \right)\chi=0.
\end{equation}
Using the explicit form of the FLRW metric and electromagnetic field, we rewrite Eq.~\eqref{Lag_Eyl} in the following form:
\begin{equation}
\label{Lag_Eyl_2}
    \chi''+2\frac{a'}{a}\chi' + \left(\frac{m^{2}}{a^2}+6\xi \frac{a''}{a}\right)\chi - (\boldsymbol{\partial}-ie\boldsymbol{A})^2\chi=0.
\end{equation}

Promoting the scalar field $\chi$ to the corresponding quantum operator, we expand it over the set of creation and annihilation operators of particles $(\hat{b}^{\dagger}_{\boldsymbol{k}},\, \hat{b}_{\boldsymbol{k}})$ and antiparticles $(\hat{c}^{\dagger}_{\boldsymbol{k}},\, \hat{c}_{\boldsymbol{k}})$ with different momenta $\boldsymbol{k}$:
\begin{equation}
\label{Rozklad}
    \hat{\chi} \left(\eta,\boldsymbol{x} \right)=\int\frac{d^{3}\boldsymbol{k}}{(2\pi)^{3/2}a(\eta)}\left[\hat{b}_{\boldsymbol{k}}\chi_{\boldsymbol{k}}(\eta) e^{i\boldsymbol{k}\cdot\boldsymbol{x}} + \hat{c}^{\dagger}_{\boldsymbol{k}} \chi^{*}_{-\boldsymbol{k}}(\eta) e^{-i\boldsymbol{k}\cdot\boldsymbol{x}} \right]\,.
\end{equation}
The annihilation and creation operators satisfy the canonical commutation relations among which the only nontrivial ones are
\begin{equation}
\label{komutatory}
[\hat{b}_{\boldsymbol{k}}, \hat{b}^{\dagger}_{\boldsymbol{k}'}] = [\hat{c}_{\boldsymbol{k}}, \hat{c}^{\dagger}_{\boldsymbol{k}'}]  = \delta^{(3)}(\boldsymbol{k}-\boldsymbol{k}')\,.
\end{equation}
The factor $a(\eta)$ in denominator in Eq.~\eqref{Rozklad} was introduced for further convenience.

Inserting the expansion \eqref{Rozklad} into the equation of motion \eqref{Lag_Eyl_2} we get the mode equation governing the evolution of the mode function $\chi_{\boldsymbol{k}}(\eta)$:
\begin{equation}
\label{Oscyl}
    \chi''_{\boldsymbol{k}}(\eta)+\Omega_{\boldsymbol{k}}^{2}(\eta)\chi_{\boldsymbol{k}}(\eta)=0\,.
\end{equation}
It has an oscillatorlike form with the time-dependent effective frequency
\begin{equation}
\label{Omega}
    \Omega_{\boldsymbol{k}}^{2}(\eta)=(\boldsymbol{k}-e\boldsymbol{A})^{2}+m^{2}a^{2}+(6\xi-1)\frac{a''}{a}.
\end{equation}
It is easy to observe from Eq.~\eqref{Omega} that in the absence of an electromagnetic field in a nonexpanding universe, the frequency $\Omega_{\boldsymbol{k}}$ does not depend on time $\eta$. Then, Eq.~\eqref{Oscyl} has two independent solutions which describe the positive- and negative-frequency modes. If the initial condition imposed at a certain moment of time corresponds, e.g., to a positive-frequency mode, the solution remains to be positive-frequency all the time. Therefore, we conclude that the creation of particles in Minkowski spacetime does not occur.

Moreover, even in the expanding universe, for the massless scalar particle, $m=0$, with the conformal nonminimal coupling to gravity, $\xi=1/6$, and in the absence of the electric field, equation of motion \eqref{Oscyl} has a very simple oscillatorlike form  $\chi''_{\boldsymbol{k}}+k^2 \chi_{\boldsymbol{k}}=0$ whose positive-frequency solution is a well-known Bunch--Davies vacuum~\cite{Bunch:1978yq}:
\begin{equation}
\label{chi-BD}
    \chi_{\boldsymbol{k},+}^{\mathrm{BD}}(\eta)=\frac{1}{\sqrt{2k}} e^{-ik\eta}\, .
\end{equation}
Being prepared in this state initially, the scalar field will remain in it forever meaning that there is no particle production even in the expanding universe. This is a consequence of the conformal invariance of the action of a massless conformally-coupled scalar field~\cite{Parker:1969au}.
Here, we would like to note that this conformal symmetry becomes explicit only if we are working in the FLRW metric written in terms of the conformal time which is conformally flat.

The situation changes in the presence of a nonzero electric field in the expanding universe. Indeed, the value of $\Omega_{\boldsymbol{k}}$ becomes dependent on the conformal time $\eta$, and therefore, in general, it is impossible to find exact general solution to the mode equation~\eqref{Oscyl} and separate positive- and negative-frequency modes\footnote{Note, that in certain particular cases when the time dependences of the scale factor and the electric field have a simple form, one can find exact analytical solution to Eq.~\eqref{Oscyl}; see, e.g., Refs.~\cite{Kobayashi:2014zza,Frob:2014zka,Bavarsad:2016cxh} for the case of a minimally coupled complex scalar field in the constant electric field in de Sitter spacetime.}. Nevertheless, one can still construct an approximate solution by employing the Wentzel--Kramers--Brillouin approximation up to a certain adiabatic order. For example, in the zeroth adiabatic order, the positive-frequency solution has the form
\begin{equation}
\label{chi-0-adiabat}
    \chi_{\boldsymbol{k},+}^{(0)}(\eta)=\frac{1}{\sqrt{2\Omega_{\boldsymbol{k}}(\eta)}} e^{-i\int^{\eta}\Omega_{\boldsymbol{k}}(\eta')d\eta'\,}\, ,
\end{equation}
where the phase of this function can be fixed by choosing the lower integration limit in the exponent. For modes with very large momenta this expression reduces to Eq.~\eqref{chi-BD} for the Bunch--Davies vacuum. Now, let us require that at a given moment of time $\eta_0$ the exact mode function $\chi_{\boldsymbol{k}}$ matches the one in Eq.~\eqref{chi-0-adiabat}. Then, the annihilation and creation operators appearing in decomposition~\eqref{Rozklad} with such mode functions define the adiabatic vacuum (of zeroth order). Consequently, at the moment of time $\eta_0$ there will be no particles in the system. However, at any later moment of time, the exact mode function does not match Eq.~\eqref{chi-0-adiabat} and in general represents a mixture of positive- and negative-frequency adiabatic modes. In such a case, it is convenient to describe it by the Bogolyubov coefficients $\alpha_{\boldsymbol{k}}$ and $\beta_{\boldsymbol{k}}$ in the following way:
\begin{equation}
\label{chi_k_Bogolubov}
    \chi_{\boldsymbol{k}}(\eta) = \frac{1}{\sqrt{2\Omega_{\boldsymbol{k}}(\eta)}} \left[\alpha_{\boldsymbol{k}}(\eta) e^{-i\Theta_{\boldsymbol{k}}(\eta)} + \beta_{\boldsymbol{k}}(\eta) e^{i\Theta_{\boldsymbol{k}}(\eta)} \right]\, , 
\end{equation}
where  
\begin{equation}
    \Theta_{\boldsymbol{k}}(\eta)=\int_{\eta_{0}}^{\eta}\Omega_{\boldsymbol{k}}(\eta')d\eta'\, .
\end{equation}
The Bogolyubov coefficients satisfy the normalization condition
\begin{equation}
\label{Bogolubov}
    \lvert\alpha_{\boldsymbol{k}}(\eta)\rvert^{2} - \lvert\beta_{\boldsymbol{k}}(\eta)\rvert^{2} = 1\, ,
\end{equation}
and at the moment of time $\eta_0$ they satisfy the initial conditions $\alpha_{\boldsymbol{k}}(\eta_{0})=1$ and $\beta_{\boldsymbol{k}}(\eta_{0})=0$ (by construction of the adiabatic vacuum at the moment $\eta_0$).

It is straightforward to show that the mode equation~\eqref{Oscyl} is identically satisfied if the evolution of the Bogolyubov coefficients is determined by the following system of equations:
\begin{subequations}
\label{eq_Bogolubov}
\begin{align}
    \alpha_{\boldsymbol{k}}' &= \dfrac{\Omega_{\boldsymbol{k}}'}{2\Omega_{\boldsymbol{k}}}e^{2i\Theta_{\boldsymbol{k}}}\beta_{\boldsymbol{k}}\, ,\\
    \beta_{\boldsymbol{k}}' &= \dfrac{\Omega_{\boldsymbol{k}}'}{2\Omega_{\boldsymbol{k}}}e^{-2i\Theta_{\boldsymbol{k}}}\alpha_{\boldsymbol{k}}\, .
\end{align}
\end{subequations}

Taking into account the normalization condition in Eq.~\eqref{Bogolubov}, we conclude that two complex functions $\alpha_{\boldsymbol{k}}$ and $\beta_{\boldsymbol{k}}$ have only three independent real degrees of freedom, which can be conveniently parameterized as follows:
\begin{align}
\label{F_k}
    \mathcal{F}_{\boldsymbol{k}}(\eta) &= \lvert\beta_{\boldsymbol{k}}(\eta)\rvert^{2}\, ,\\
\label{G_k}
    \mathcal{G}_{\boldsymbol{k}}(\eta) &= \operatorname{Re}\left(\alpha_{\boldsymbol{k}}\,\beta^{*}_{\boldsymbol{k}}\,e^{-2i\Theta_{\boldsymbol{k}}(\eta)}\right)\, ,\\
\label{H_k}
    \mathcal{H}_{\boldsymbol{k}}(\eta) &= \operatorname{Im}\left(\alpha_{\boldsymbol{k}}\,\beta^{*}_{\boldsymbol{k}}\,e^{-2i\Theta_{\boldsymbol{k}}(\eta)} \right)\, .
\end{align}
Then, using Eqs.~\eqref{eq_Bogolubov}, one can derive equations of motion for quantities \eqref{F_k}--\eqref{H_k}:
\begin{subequations}
\label{system_FGH}
    \begin{align}
        \mathcal{F}_{\boldsymbol{k}}'(\eta) & = \frac{\Omega_{\boldsymbol{k}}'}{\Omega_{\boldsymbol{k}}}\,\mathcal{G}_{\boldsymbol{k}}(\eta)\, ,\\
        \mathcal{G}_{\boldsymbol{k}}'(\eta) & = \frac{\Omega_{\boldsymbol{k}}'}{2\Omega_{\boldsymbol{k}}}\big[1 + 2\mathcal{F}_{\boldsymbol{k}}(\eta)\big] +2\Omega_{\boldsymbol{k}}\mathcal{H}_{\boldsymbol{k}}(\eta)\, ,\\
        \mathcal{H}_{\boldsymbol{k}}'(\eta) & = -2\Omega_{\boldsymbol{k}} \,\mathcal{G}_{\boldsymbol{k}}(\eta)\, .
    \end{align}
\end{subequations}
The main advantages of this system of equations compared to \eqref{eq_Bogolubov} is that it is real and does not contain fast oscillating coefficients. In order to rewrite it in the final form, we switch from the canonical momentum $\boldsymbol{k}$ to the kinetic momentum in the conformal spacetime $\boldsymbol{p}=\boldsymbol{k}-e\boldsymbol{A}$ and introduce the corresponding energy $\epsilon_{\boldsymbol{p}}\equiv \sqrt{\boldsymbol{p}^{2}+m^{2}a^{2}}$.
\footnote{The quantities $\boldsymbol{p}$ and $\epsilon_{\boldsymbol{p}}$ do not coincide with the physical kinetic momentum and energy of the scalar particle measured by the comoving cosmological observer. They are introduced for convenience in further analysis. The corresponding physical quantities can be expressed as $\boldsymbol{p}_{\mathrm{phys}}=\boldsymbol{p}/a$ and $\epsilon_{\boldsymbol{p},\mathrm{phys}}=\epsilon_{\boldsymbol{p}}/a$.}
Further, we perform the following replacements in the system of equations~\eqref{system_FGH}:
\begin{equation}
\label{omega_p}
    \Omega_{\boldsymbol{k}}\, \to\, \omega(\eta,\boldsymbol{p})=\sqrt{\epsilon_{\boldsymbol{p}}^{2}+(6\xi-1)\frac{a''}{a}}\, ,
\end{equation}
\begin{equation}
\label{FGH}
    \mathcal{F}_{\boldsymbol{k}}(\eta)\, \to\, \mathcal{F}(\eta, \boldsymbol{p})\, , \qquad
    \mathcal{G}_{\boldsymbol{k}}(\eta)\, \to\, \mathcal{G}(\eta, \boldsymbol{p})\, , \qquad
    \mathcal{H}_{\boldsymbol{k}}(\eta)\, \to\, \mathcal{H}(\eta, \boldsymbol{p})\, ,
\end{equation}
\begin{equation}
\label{Q}
    \frac{\Omega_{\boldsymbol{k}}'}{\Omega_{\boldsymbol{k}}}\, \to\,  Q(\eta,\boldsymbol{p})\equiv \frac{1}{\omega(\eta,\boldsymbol{p})^{2}}\Bigg[e\,a^{2}(\boldsymbol{p}\cdot\boldsymbol{E} ) + m^{2}aa'+\frac{6\xi-1}{2}\left(\frac{a'''}{a}-\frac{a'a''}{a^{2}} \right)  \Bigg]\, .
\end{equation}
Further, we would like to note that the time derivative is changed according to the rule
\begin{equation}
\label{d_F}
    \mathcal{F}'_{\boldsymbol{k}}(\eta)\,\to\, \frac{d}{d\eta}\mathcal{F}(\eta, \boldsymbol{p})=\left[\frac{\partial}{\partial\eta}+\frac{\partial\boldsymbol{p}}{\partial\eta}\!\cdot\!\frac{\partial}{\partial\boldsymbol{p}} \right]\!\mathcal{F}(\eta, \boldsymbol{p})= \left[\frac{\partial}{\partial\eta}+e\,a^{2}\boldsymbol{E}\!\cdot\!\frac{\partial}{\partial\boldsymbol{p}} \right] \mathcal{F}(\eta, \boldsymbol{p})\,  
\end{equation}
and similar relations also hold for the quantities $\mathcal{G}(\eta, \boldsymbol{p}), \mathcal{H}(\eta, \boldsymbol{p})$.

Finally, we obtain the system of quantum Vlasov equations for the kinetic functions \eqref{FGH}:
\begin{subequations}
\label{eq_F_G_H}
\begin{align}
    \left(\frac{\partial}{\partial\eta}+e\,a^{2}\boldsymbol{E}\!\cdot\!\frac{\partial}{\partial\boldsymbol{p}} \right) \mathcal{F}(\eta, \boldsymbol{p}) &= Q(\eta,\boldsymbol{p}) \mathcal{G}(\eta, \boldsymbol{p})\, , \\
    \left(\frac{\partial}{\partial\eta}+e\,a^{2}\boldsymbol{E}\!\cdot\!\frac{\partial}{\partial\boldsymbol{p}} \right) \mathcal{G}(\eta, \boldsymbol{p})\, &= \frac{1}{2} Q(\eta,\boldsymbol{p})\left[1+2\mathcal{F}(\eta, \boldsymbol{p}) \right] + 2\omega(\eta,\boldsymbol{p})\mathcal{H}(\eta, \boldsymbol{p})\, ,\\
    \left(\frac{\partial}{\partial\eta}+e\,a^{2}\boldsymbol{E}\!\cdot\!\frac{\partial}{\partial\boldsymbol{p}} \right) \mathcal{H}(\eta, \boldsymbol{p}) &= -2\omega(\eta,\boldsymbol{p})\mathcal{G}(\eta, \boldsymbol{p})\, .
\end{align}    
\end{subequations}
These equations describe the creation of charged scalar particles in the expanding universe in the presence of the spatially homogeneous electric field. We would like to mention an important property of this system of equations: it ensures that the combination $(1+2\mathcal{F})^2-(2\mathcal{G})^2-(2\mathcal{H})^2$ is conserved in time and equals to unity. This can also be seen directly from the definitions in Eqs.~\eqref{F_k}--\eqref{H_k} together with normalization condition \eqref{Bogolubov}. According to Ref.~\cite{Kim:2011jw}, this is a manifestation of the charge conservation and is related to the invariance of the Casimir operator for the SU(1,1) algebra~\cite{Perelomov:1972sro}.

\section{Physical observables}
\label{sec:observables}

In this section we determine the main physical observables for the scalar field, namely the electric current and energy--momentum tensor, and express them in terms of kinetic functions $\mathcal{F}$, $\mathcal{G}$, and $\mathcal{H}$ introduced in the previous section.

\subsection{Electric current}
\label{subsec:current-def}

We start with the electric current which is the vacuum expectation value of the corresponding quantum operator:
\begin{equation}
    j^{\mu}=\langle \frac{\partial \mathcal{L}_{\mathrm{ch}}}{\partial A_{\mu}}\rangle = (0,\,\frac{1}{a^2}\boldsymbol{j})\, .
\end{equation}
Using the explicit form of the scalar-field Lagrangian density in Eq.~\eqref{L_ch}, we get
\begin{equation}
\label{j}
    \boldsymbol{j}=ie\langle\hat{\chi}^{\dagger}\,\boldsymbol{\nabla}\hat{\chi}-\hat{\chi}\,\boldsymbol{\nabla}\hat{\chi}^{\dagger}-2ie\boldsymbol{A}\,\hat{\chi}^{\dagger}\hat{\chi} \rangle\, . 
\end{equation}
Further, we substitute the operator decomposition \eqref{Rozklad} and express the electric current in terms of the mode function:
\begin{equation}
\label{j2}
    \boldsymbol{j}=-\frac{2e}{a^{2}}\!\int\!\!\frac{d^{3}\boldsymbol{k}}{(2\pi)^{3}}(\boldsymbol{k}-e\boldsymbol{A})\,\lvert\chi_{\boldsymbol{k}}(\eta)\rvert^{2}\, .
\end{equation}

Finally, we use Eqs.~\eqref{chi_k_Bogolubov}, \eqref{F_k}--\eqref{H_k} and rewrite Eq.~\eqref{j2} in terms of the kinetic functions $\mathcal{F}$, $\mathcal{G}$, and  $\mathcal{H}$ as follows:
\begin{equation}
\label{j3}
    \boldsymbol{j}=-\frac{2e}{a^{2}}\!\int\!\!\frac{d^{3}\boldsymbol{k}}{(2\pi)^{3}}(\boldsymbol{k}-e\boldsymbol{A}) \frac{1+2\mathcal{F}_{\boldsymbol{k}}+2\mathcal{G}_{\boldsymbol{k}}}{2\Omega_{\boldsymbol{k}}}
    =-\frac{2e}{a^{2}}\!\int\!\frac{d^{3}\boldsymbol{p}}{(2\pi)^{3}}\,\boldsymbol{p}\,\frac{\mathcal{F}(\eta,\boldsymbol{p})+\mathcal{G}(\eta,\boldsymbol{p})}{\omega(\eta,\boldsymbol{p})}\, ,
\end{equation}
where the term with a unity in the numerator was omitted since it is odd under the change $\boldsymbol{p}\to -\boldsymbol{p}$ and, therefore, identically vanishes after integration over all momenta. 

\subsection{Energy--momentum tensor}
\label{subsec:EMT}

The energy--momentum tensor of a scalar charged field $\chi$ can be found by varying the action with respect to the spacetime metric:
\begin{align}
\label{TEI_chi}
    T_{\mu\nu}^{\chi}&=\frac{2}{\sqrt{-g}}\langle\frac{\delta S_{\chi}}{\delta g^{\mu\nu}} \rangle \nonumber\\
    &=\langle  (\mathcal{D}_{\mu}\hat{\chi})^{\dagger}(\mathcal{D}_{\nu}\hat{\chi})+(\mathcal{D}_{\nu}\hat{\chi})^{\dagger}(\mathcal{D}_{\mu}\hat{\chi}) -  g_{\mu\nu}[(\mathcal{D}_{\alpha}\hat{\chi})^{\dagger}(\mathcal{D}^{\alpha}\hat{\chi})-m^{2}\hat{\chi}^{\dagger}\hat{\chi}]\rangle \nonumber \\
    &+ 2\xi \langle \Big(R_{\mu\nu}-\frac{1}{2}R g_{\mu\nu} - \nabla_{\mu}\,\nabla_{\nu} + g_{\mu\nu}\,\nabla_{\alpha}\,\nabla^{\alpha}\Big)\hat{\chi}^{\dagger}\hat{\chi}\rangle\, .
\end{align}

Then, we immediately have the energy density of the produced particles $\rho_{\chi} \equiv T_{\,\,0}^{\chi\,0}$:
\begin{align}
\label{rho}
    \rho_{\chi} &= \frac{1}{a^2}\langle (\mathcal{D}_{0}\hat{\chi})^{\dagger}(\mathcal{D}_{0}\hat{\chi})+(\mathcal{D}_{i}\hat{\chi})^{\dagger}(\mathcal{D}_{i}\hat{\chi})+ \Big(m^{2}a^2 + 6\xi\frac{a^{\prime 2}}{a^{2}} \Big)\hat{\chi}^{\dagger}\hat{\chi}\rangle \nonumber\\
    &-\frac{2\xi}{a^2}\langle\Big(\partial_{i}^{2} - 3\frac{a'}{a}\partial_{0}\Big)  \hat{\chi}^{\dagger}\hat{\chi}\rangle\, .
\end{align}
Instead of pressure, $P_{\chi}=(1/3)g^{ij}T_{ij}^{\chi}$, it is more convenient to work with the trace of the energy--momentum tensor $T_{\chi}=g^{\mu\nu}T_{\mu\nu}^{\chi}$ which has the form:
\begin{equation}
\label{T_chi}
    T_{\chi}=\langle 2(6\xi-1)\left[(\mathcal{D}_{\alpha}\hat{\chi})^{\dagger}(\mathcal{D}^{\alpha}\hat{\chi})+\xi R \hat{\chi}^{\dagger}\hat{\chi} \right] + 4(1-3\xi)m^{2}\hat{\chi}^{\dagger}\hat{\chi}\rangle\, .
\end{equation}

Again, using the decomposition of $\hat{\chi}$ over annihilation and creation operators in Eq.~\eqref{Rozklad}, we express the energy density and trace in terms of the mode function:
\begin{equation}
\label{rho2}
    \rho_{\chi}=\!\int\!\!\frac{d^{3}\boldsymbol{k}}{(2\pi)^{3}\,a^{4}}\Big\{ \Big|\chi_{\boldsymbol{k}}'-\frac{a'}{a}\chi_{\boldsymbol{k}} \Big|^{2}+ \Big[m^{2}a^2-6\xi\frac{a^{\prime 2}}{a^{2}}+(\boldsymbol{k}-e\boldsymbol{A})^{2} + 6\xi\frac{a'}{a}\partial_{0}\Big]\big|\chi_{\boldsymbol{k}}\big|^{2}\Big\}\, ,
\end{equation}
\begin{equation}
\label{T_chi_2}
    T_{\chi}=\!\int\!\!\frac{d^{3}\boldsymbol{k}}{(2\pi)^{3}\,a^{4}}\Big\{2(6\xi-1)\Big[\Big|\chi_{\boldsymbol{k}}'-\frac{a'}{a}\chi_{\boldsymbol{k}} \Big|^{2}-\Big(\Omega_{\boldsymbol{k}}^{2}+\frac{a''}{a}\Big) \big|\chi_{\boldsymbol{k}}\big|^{2}  \Big] + 2m^{2}a^2\big|\chi_{\boldsymbol{k}}\big|^{2} \Big\}\, .
\end{equation}

Finally, expressing them in terms of the kinetic functions $\mathcal{F}$, $\mathcal{G}$, and $\mathcal{H}$ we obtain
\begin{align}
\label{rho3}
    \rho_{\chi}&=\int\frac{d^{3}\boldsymbol{p}}{(2\pi)^{3}\,a^{4}}\Big\{ \omega(\eta,\boldsymbol{p})\big[1+2\mathcal{F}(\eta,\boldsymbol{p})\big]+2\frac{a'}{a}(6\xi-1)\mathcal{H}(\eta,\boldsymbol{p})\nonumber\\
    & -\frac{6\xi-1}{\omega(\eta,\boldsymbol{p})}\Big(\frac{a^{\prime 2}}{a^{2}}+\frac{a''}{a} \Big) \Big[\frac{1}{2}+\mathcal{F}(\eta,\boldsymbol{p})+\mathcal{G}(\eta,\boldsymbol{p})\Big]  \Big\}\, ,
\end{align}
\begin{align}
\label{T_chi_3}
    T_{\chi}&=\int \frac{d^{3}\boldsymbol{p}}{(2\pi)^{3}\,a^{4}} \Big\{ - 4(6\xi-1)\Big[\omega(\eta,\boldsymbol{p})\mathcal{G}(\eta,\boldsymbol{p})+\frac{a'}{a} \mathcal{H}(\eta,\boldsymbol{p}) \Big] \nonumber\\
    &+\frac{2}{\omega(\eta,\boldsymbol{p})} \Big[m^{2}a^2-(6\xi-1) \Big(\frac{a''}{a}-\frac{a^{\prime 2}}{a^{2}} \Big)\Big]\Big[\frac{1}{2}+\mathcal{F}(\eta,\boldsymbol{p})+\mathcal{G}(\eta,\boldsymbol{p})\Big] \Big\}\, .
\end{align}

We will see below that all physical observables derived in this section contain UV divergences and, therefore, must be renormalized in order to obtain a physical meaning. We perform the renormalization in the following section.

\section{Renormalization}
\label{sec:renormalization}

In this section, we study the behavior of the kinetic functions at large values of momentum and show that the integrals in Eqs.~\eqref{j3}, \eqref{rho3}, and \eqref{T_chi_3} are divergent in the UV region. Further, we extract the divergent parts using the dimensional regularization and introduce the counterterms in the action which cancel those divergences.

\subsection{Asymptotical behavior of the kinetic functions at high energies}
\label{subsec:asyptotics}

The coefficients $\omega(\eta,\boldsymbol{p})$ and $Q(\eta,\boldsymbol{p})$ in the quantum Vlasov equations \eqref{eq_F_G_H} in the limit of large momenta, $p\equiv |\boldsymbol{p}|\!\to\!\infty$, behave as
\begin{equation}
\label{asympt-coef}
    \omega(\eta,\boldsymbol{p})=p+\mathcal{O}(p^{-1})\, , \qquad Q(\eta,\boldsymbol{p})=ea^2 (\hat{\boldsymbol{p}}\cdot\boldsymbol{E}) \, p^{-1}+\mathcal{O}(p^{-2})\, 
\end{equation}
where $\hat{\boldsymbol{p}}=\boldsymbol{p}/p$. Then, from Eq.~\eqref{eq_F_G_H}, one can see that at high momenta, $\mathcal{F}\propto p^{-4}$, $\mathcal{G}\propto p^{-3}$, and $\mathcal{H}\propto p^{-2}$, i.e., the kinetic functions demonstrate power-law behavior which may potentially lead (and indeed leads as we will see below) to the UV divergences in the observables. In order to resolve the latter issue, we need the exact expressions for the first few terms of the large-$p$ expansion of the kinetic functions. For the technical details, we refer the reader to Appendix~\ref{app-expansion} while here we just list the results:
\begin{align}
    \mathcal{F}(\eta,\boldsymbol{p})&=\frac{e^{2}a^{4}(\boldsymbol{v}\cdot\boldsymbol{E})^{2}}{16\,\epsilon_{\boldsymbol{p}}^{4}} + \mathcal{O}(\epsilon_{\boldsymbol{p}}^{-5})\, ,\label{F-asym} \\
    \mathcal{G}(\eta,\boldsymbol{p})&=\frac{e\,a^2}{8\,\epsilon_{\boldsymbol{p}}^{3}}\, \boldsymbol{v}\cdot\Big(\boldsymbol{E}'+2\frac{a'}{a}\boldsymbol{E} \Big) + \frac{e^{2}a^{4}\big[\boldsymbol{E}^{2}-3(\boldsymbol{v}\cdot\boldsymbol{E})^2\big]}{8\,\epsilon_{\boldsymbol{p}}^{4}} \label{G-asym}\\
    &+ \frac{1}{8\,\epsilon_{\boldsymbol{p}}^{4}}\!\bigg[m^{2}(a'^{\,2}+aa'') + (6\xi-1)\left(\frac{a^{\mathrm{IV}}}{2a}-\frac{a''^{\,2}}{2a^{2}}-\frac{a'a'''}{a^{2}} + \frac{a'^{\,2}a''}{a^{3}} \right)  \bigg]+\mathcal{O}(\epsilon_{\boldsymbol{p}}^{-5})\, ,\nonumber\\ 
    \mathcal{H}(\eta,\boldsymbol{p})&=-\frac{e\,a^{2}\,(\boldsymbol{v}\cdot\boldsymbol{E})}{4\,\epsilon_{\boldsymbol{p}}^{2}}-\frac{1}{4\, \epsilon_{\boldsymbol{p}}^{3}}\!\left[m^{2}aa'\!+\!\frac{6\xi-1}{2}\left(\frac{a'''}{a}-\frac{a'a''}{a^{2}} \right)  \right]+\mathcal{O}(\epsilon_{\boldsymbol{p}}^{-4})\, , \label{H-asym}
\end{align}
where $\boldsymbol{v}=\boldsymbol{p}/\epsilon_{\boldsymbol{p}}$ is the velocity of the particle and we presented only those terms which may lead to a divergent contribution to observables. Note that the kinetic functions are expressed in terms of inverse powers of $\epsilon_{\boldsymbol{p}}$ rather than $p$. Among the two expansions which are completely equivalent in the UV region the expansion in inverse powers of $\epsilon_{\boldsymbol{p}}$ is more convenient since it does not cause the problems in the infrared limit $p\to 0$.

\subsection{Electric current}
\label{subsec:renorm-current}

Using Eq.~\eqref{j3} for the electric current and the asymptotical expansions for the kinetic functions in Eqs.~\eqref{F-asym}--\eqref{H-asym}, it is easy to see that (i)~the integral over the momentum is logarithmically divergent in the UV region and (ii)~the divergence comes only from the term in $\mathcal{G}(\eta,\boldsymbol{p}$ which behaves as $\propto \epsilon_{\boldsymbol{p}}^{-3}$, i.e, from the first term in Eq.~\eqref{G-asym}.
Then, the $i$th component of the electric current reads as:
\begin{equation}
\label{j-exp}
j^{i}=-\frac{e^{2}}{4}\int\frac{d^{3}\boldsymbol{p}}{(2\pi)^{3}}\,\left[\frac{v^i v^j}{\epsilon_{\boldsymbol{p}}^3}\,\left( E^{j \prime}+2\frac{a'}{a}E^j \right) + \mathcal{O}(\epsilon_{\boldsymbol{p}}^{-4})\right]\, , 
\end{equation}
where we wrote explicitly only the term which leads to the divergence.

In order to extract the divergent contribution, we employ the dimensional regularization switching from the (3+1)-dimensional space time to a general $d$-dimensional one. Then, for the divergent integral in Eq.~\eqref{j-exp} by switching to the physical momentum $\boldsymbol{q}=\boldsymbol{p}/a$ we get:
\begin{align}
    \int\frac{d^{3}\boldsymbol{p}}{(2\pi)^{3}}\,\frac{v^i v^j}{\epsilon_{\boldsymbol{p}}^3}&= \int\frac{d^{3}\boldsymbol{q}}{(2\pi)^{3}}\,\frac{v^i v^j}{(q^2+m^2)^{3/2}} \to \mu_{\mathrm{r}}^{4-d}\int\frac{d^{d-1}\boldsymbol{q}}{(2\pi)^{d-1}}\,\frac{v^i v^j}{(q^2+m^2)^{3/2}} \nonumber\\
    &= \frac{\delta^{ij}\Omega_{(d-1)}\, \mu_{\mathrm{r}}^{4-d}}{(d-1)(2\pi)^{d-1}} \!\!\int\limits_{0}^{+\infty}\!\! dq\, q^{d} (q^2+m^2)^{-\frac52} = \frac{\delta^{ij}}{12\pi^2} \Big(\frac{m^2}{4\pi\mu_{\mathrm{r}}^2}\Big)^{\frac{d-4}{2}}\Gamma\Big(\frac{4-d}{2}\Big)\nonumber\\
    &=\frac{\delta^{ij}}{12\pi^2}\Big[\frac{2}{4-d}-\gamma_{\mathrm{E}}-\ln\Big(\frac{m^2}{4\pi\mu_{\mathrm{r}}^2}\Big)+\mathcal{O}(4-d)\Big]\, ,\label{dimreg-1}
\end{align}
where $\Omega_{(d-1)}=2\pi^{(d-1)/2}/\Gamma(\tfrac{d-1}{2})$ is the full solid angle in $(d-1)$-dimensional space; $\mu_{\mathrm{r}}$ is a free parameter with the dimension of mass, which arises in order to compensate the mass dimension of the original integral and has a meaning of the renormalization energy scale (for simplicity, in what follows we set it $\mu_{\mathrm{r}}=m$); and $\gamma_{\mathrm{E}}\approx 0.577$ is the Euler--Mascheroni constant. Obviously, the first term in brackets in Eq.~\eqref{dimreg-1} tends to infinity in the limit $d\to 4$ which restores the correct dimension of our spacetime; therefore, it must be subtracted. However, together with this term, one can subtract any finite contributions. This ambiguity is resolved only by convention. One of the most popular subtraction schemes, the $\overline{\mathrm{MS}}$ scheme, requires to subtract the combination
\begin{equation}
    \frac{1}{\bar{\varepsilon}}\equiv \frac{2}{4-d}-\gamma_{\mathrm{E}}+\ln 4\pi\, .
\end{equation}

Then, from Eqs.~\eqref{j-exp} and \eqref{dimreg-1} we conclude that the divergent part of the electric current can be written as
\begin{equation}
\label{j_div}
    \boldsymbol{j}_{\mathrm{div}}=-\frac{e^{2}}{48\pi^{2}} \frac{1}{\bar{\varepsilon}} \left(\boldsymbol{E}'+2\frac{a'}{a}\boldsymbol{E} \right)\, .
\end{equation}

Subtracting from the full electric current in Eq.~\eqref{j3} the first term in Eq.~\eqref{j-exp}, we obtain the regular (i.e., final) part of the current:
\begin{equation}
\label{j_reg}
\boldsymbol{j}_{\mathrm{reg}}=-2e\!\int\!\!\frac{d^{3}\boldsymbol{p}}{(2\pi)^{3}}\,\boldsymbol{p}\, \left\lbrace\frac{\mathcal{F}(\eta,\boldsymbol{p})+\mathcal{G}(\eta,\boldsymbol{p})}{a^2 \omega(\eta,\boldsymbol{p})}-e\frac{\big(\boldsymbol{p}\!\cdot\![\boldsymbol{E}'+2(a'/a)\boldsymbol{E}]\big)}{8\,\epsilon_{\boldsymbol{p}}^{5}} \right\rbrace\, . 
\end{equation}

The divergent contribution to the electric current can be canceled out from the Maxwell equation by introducing the counterterm in the action of the following form:
\begin{equation}
\label{S_Z3}
    \delta S_{Z_{3}}=(Z_{3}-1)\int d^{4}x \, \sqrt{-g} \left(-\frac{1}{4}F_{\mu\nu}F^{\mu\nu} \right). 
\end{equation}
This counterterm together with Eqs.~\eqref{S} and \eqref{L_EM} implies the following Maxwell equation for the electric field:
\begin{equation}
    (1+ Z_3 - 1)\Big(\boldsymbol{E}'+2\frac{a'}{a}\boldsymbol{E}\Big) = \boldsymbol{j}_{\mathrm{div}} + \boldsymbol{j}_{\mathrm{reg}}\, .
\end{equation}
Substituting here Eq.~\eqref{j_div} and requiring that after cancellation this equation takes the renormalized form containing only finite quantities
\begin{equation}
    \boldsymbol{E}'+2\frac{a'}{a}\boldsymbol{E} = \boldsymbol{j}_{\mathrm{reg}}\, ,
\end{equation}
we obtain the value of the renormalization parameter
\begin{equation}
\label{Z3}
    Z_3 = 1 - \frac{e^{2}}{48\pi^{2}} \frac{1}{\bar{\varepsilon}}\, ,
\end{equation}
which reproduces the well-known result for the charge renormalization parameter in the scalar quantum electrodynamics in Minkowski spacetime, see, e.g., Ref.~\cite{Srednicki-book}.

\subsection{Energy density}
\label{subsec:renorm-rho}

Again, using the asymptotical expansions in Eqs.~\eqref{F-asym}--\eqref{H-asym} we expand the integral in Eq.~\eqref{rho3} keeping the terms leading to the UV divergence. We get the following expression for the energy density:
\begin{align}
\label{rho_exp}
    \rho_{\chi}&=\int\frac{d^{3}\boldsymbol{p}}{(2\pi)^{3}\,a^{4}}\bigg\{ \epsilon_{\boldsymbol{p}}-\frac{6\xi-1}{2\epsilon_{\boldsymbol{p}}} \frac{a^{\prime 2}}{a^{2}}- \frac{6\xi-1}{2\epsilon_{\boldsymbol{p}}^{2}}\frac{a'}{a}\,e(\boldsymbol{v}\!\cdot\!\boldsymbol{E})+\frac{e^{2}a^{4}(\boldsymbol{v}\!\cdot\!\boldsymbol{E})^{2}}{8\,\epsilon_{\boldsymbol{p}}^{3}} \nonumber\\
    &- \frac{6\xi-1}{2\,\epsilon_{\boldsymbol{p}}^{3}}\!\bigg[m^{2}a^{\prime 2}+\frac{6\xi-1}{2}\!\Big(\frac{a'a'''}{a^{2}}\!-\!\frac{2a^{\prime 2}a''}{a^{3}}\!-\!\frac{a^{\prime\prime 2}}{2a^{2}} \Big) \! \bigg] + \mathcal{O}(\epsilon_{\boldsymbol{p}}^{-4}) \bigg\}\, .
\end{align}
Here, the third term in curly brackets vanishes after integration since it is odd under the change $\boldsymbol{p}\to -\boldsymbol{p}$. All other terms are divergent and can be computed in the dimensional regularization by switching to a $d$-dimensional spacetime as shown in Eq.~\eqref{dimreg-1} or in a similar way:
\begin{align}
    \int\frac{d^{3}\boldsymbol{p}}{(2\pi)^{3}}\,\epsilon_{\boldsymbol{p}}&\to 
    -\frac{m^4 a^4}{4\pi^2} \Big(\frac{m^2}{4\pi\mu_{\mathrm{r}}^2}\Big)^{\frac{d-4}{2}}\frac{\Gamma\Big(\frac{4-d}{2}\Big)}{d (d-2)}=-\frac{m^4 a^4}{32\pi^2}\bigg[\frac{1}{\bar{\varepsilon}}+\frac{3}{2}+\mathcal{O}(4-d)\bigg]\, ,\label{dimreg-2} \\
    \int\frac{d^{3}\boldsymbol{p}}{(2\pi)^{3}}\,\frac{1}{\epsilon_{\boldsymbol{p}}}&\to 
    -\frac{m^2 a^2}{4\pi^2} \Big(\frac{m^2}{4\pi\mu_{\mathrm{r}}^2}\Big)^{\frac{d-4}{2}}\frac{\Gamma\Big(\frac{4-d}{2}\Big)}{d-2}=-\frac{m^2 a^2}{8\pi^2}\bigg[\frac{1}{\bar{\varepsilon}}+1+\mathcal{O}(4-d)\bigg]\, ,\label{dimreg-3} \\
    \int\frac{d^{3}\boldsymbol{p}}{(2\pi)^{3}}\,\frac{1}{\epsilon_{\boldsymbol{p}}^3}&\to 
    \frac{1}{4\pi^2} \Big(\frac{m^2}{4\pi\mu_{\mathrm{r}}^2}\Big)^{\frac{d-4}{2}}\Gamma\Big(\frac{4-d}{2}\Big)=\frac{1}{4\pi^2}\bigg[\frac{1}{\bar{\varepsilon}}+\mathcal{O}(4-d)\bigg]\, .\label{dimreg-4}
\end{align}

Collecting all terms proportional to $1/\bar{\varepsilon}$, we get the divergent part of the energy density:
\begin{equation}
\label{rho_div}
    \rho_{\chi}^{\mathrm{div}}=\bigg[\frac{e^{2}\boldsymbol{E}^{2}}{96\pi^{2}}-\frac{m^{4}}{32\pi^{2}} -\frac{6\xi-1}{16\pi^{2}} \frac{m^{2}a^{\prime 2}}{a^{4}} - \frac{(6\xi-1)^{2}}{16\pi^{2}}\Big(\frac{a'a'''}{a^{6}}-\frac{2a^{\prime 2}a''}{a^{7}}-\frac{a^{\prime\prime 2}}{2a^{6}} \Big)  \bigg] \frac{1}{\bar{\varepsilon}}\, .
\end{equation}

This divergent contribution can be eliminated by introducing the corresponding counterterms into the action. First of all, we note that the first term in Eq.~\eqref{rho_div}, the only one which depends on the electric field, is fully canceled by the counterterm~\eqref{S_Z3} considered in the previous subsection. The rest of the divergent terms does not depend on the electric field but only on the particle's mass and time derivatives of the scale factor. Therefore, they represent the radiative corrections to the Einstein--Hilbert action for gravity. Let us search for the counterterms which cancel those divergences using the following Anzatz:
\begin{equation}
\label{dS-grav}
    \delta S_{\mathrm{grav}}=\int d^4 x \sqrt{-g}(a_1 + a_2 R + a_3 R^2)\, .
\end{equation}
The first term, $a_1$, renormalizes the cosmological constant, $a_2$ takes care of the Planck mass, and $a_3$ renormalizes the coefficient of the Starobinsky $R^2$ term~\cite{Starobinsky:1980te}. The effective energy--momentum tensor which follows from such counterterms in the action can be obtained by varying Eq.~\eqref{dS-grav} with respect to metric:
\begin{align}
\label{T_ct}
    \delta T^{\mu\nu}_{\mathrm{grav}}=\frac{2}{\sqrt{-g}}\frac{\delta(\delta S_{\mathrm{grav}})}{\delta g_{\mu\nu}}=&-a_{1}g^{\mu\nu}+a_{2}(2R^{\mu\nu}-Rg^{\mu\nu})\nonumber\\
    &+4a_{3}\Big[RR^{\mu\nu}-\frac{1}{4}R^{2}g^{\mu\nu}-(\nabla^{\mu}\nabla^{\nu}-g^{\mu\nu}\nabla_{\alpha}\nabla^{\alpha} )R \Big]\, . 
\end{align}
The 00 component of this expression gives the corresponding energy density:
\begin{equation}
\label{rho-ct}
    \delta\rho_{\mathrm{grav}} = -a_1 + 6a_2 \frac{a^{\prime 2}}{a^{4}} - 72 a_3 \Big(\frac{a'a'''}{a^{6}}-\frac{2a^{\prime 2}a''}{a^{7}}-\frac{a^{\prime\prime 2}}{2a^{6}} \Big)\, .
\end{equation}
Now, requiring $\rho_{\chi}^{\mathrm{div}} + \delta\rho_{Z_3}+ \delta\rho_{\mathrm{grav}} = 0$, we obtain the coefficients $a_{1}$, $a_{2}$, and $a_{3}$ in the following form:
\begin{align}
\label{a1}
    a_{1}&=-\frac{m^{4}}{32\pi^{2}}\,\frac{1}{\bar{\varepsilon}}\, ,\\
\label{a2}
    a_{2}&=\left(\xi-\frac{1}{6} \right) \frac{m^{2}}{16\pi^{2}}\,\frac{1}{\bar{\varepsilon}}\, ,\\
\label{a3}
    a_{3}&=-\left(\xi-\frac{1}{6} \right)^{2}\frac{1}{32\pi^{2}}\,\frac{1}{\bar{\varepsilon}}\, ,
\end{align}
which fully fix the counterterm in Eq.~\eqref{dS-grav}.

Finally, subtracting from the integrand in Eq.~\eqref{rho3} the terms shown in Eq.~\eqref{rho_exp} and collecting the finite contributions in Eqs.~\eqref{dimreg-2}--\eqref{dimreg-3}, we obtain the expression for the regular part of the energy density:
\begin{align}
\label{rho_reg}
    \rho_{\chi}^{\mathrm{reg}}&=-\frac{3m^{4}}{64\pi^{2}}+\frac{6\xi-1}{16\pi^{2}} \frac{m^{2}a^{\prime 2}}{a^{4}}+\int\frac{d^{3}\boldsymbol{p}}{(2\pi)^{3}\,a^{4}}\bigg\{\omega(\eta,\boldsymbol{p})\big[1+2\mathcal{F}(\eta,\boldsymbol{p})\big]\nonumber\\
    &+2\frac{a'}{a}(6\xi-1)\mathcal{H}(\eta,\boldsymbol{p})-\frac{6\xi-1}{\omega(\eta,\boldsymbol{p})}\Big(\frac{a^{\prime 2}}{a^{2}}+\frac{a''}{a} \Big) \Big[\frac{1}{2}+\mathcal{F}(\eta,\boldsymbol{p})+\mathcal{G}(\eta,\boldsymbol{p})\Big]  \nonumber\\
    &-\epsilon_{\boldsymbol{p}}+\frac{6\xi-1}{2\epsilon_{\boldsymbol{p}}} \frac{a^{\prime 2}}{a^{2}}+ \frac{6\xi-1}{2\epsilon_{\boldsymbol{p}}^{2}}\frac{a'}{a}\,e(\boldsymbol{v}\!\cdot\!\boldsymbol{E})-\frac{e^{2}a^{4}(\boldsymbol{v}\!\cdot\!\boldsymbol{E})^{2}}{8\,\epsilon_{\boldsymbol{p}}^{3}} \nonumber\\
    &+ \frac{6\xi-1}{2\,\epsilon_{\boldsymbol{p}}^{3}}\!\bigg[m^{2}a^{\prime 2}+\frac{6\xi-1}{2}\!\Big(\frac{a'a'''}{a^{2}}\!-\!\frac{2a^{\prime 2}a''}{a^{3}}\!-\!\frac{a^{\prime\prime 2}}{2a^{2}} \Big)\bigg]\bigg\} \, .
\end{align}
This expression represents the physically meaningful and finite energy density of the produced scalar particles which can be used, in particular, in the Friedmann equation describing the expansion rate of the universe.

\subsection{Trace of the energy--momentum tensor}
\label{subsec:renorm-trace}

Finally, let us now consider the UV divergences in the trace of the energy--momentum tensor of the produced scalar particles. Expanding the integrand in Eq.~\eqref{T_chi_3} by using Eqs.~\eqref{F-asym}--\eqref{H-asym}, we collect the terms which might lead to the divergence:
\begin{align}
\label{T_chi_exp}
    T_{\chi}&=\int\frac{d^{3}\boldsymbol{p}}{(2\pi)^{3}\,a^{2}}\bigg\{- e\frac{a'}{a}\frac{(\boldsymbol{v}\cdot\boldsymbol{E})}{4\epsilon_{\boldsymbol{p}}^{2}} -\frac{(6\xi-1)e}{2\epsilon_{\boldsymbol{p}}^{2}} \boldsymbol{v}\cdot\Big(\boldsymbol{E}'+2\frac{a'}{a}\boldsymbol{E}\Big)\nonumber\\
    &-\frac{(6\xi-1)a^2 e^2}{2\epsilon_{\boldsymbol{p}}^{3}}\big[\boldsymbol{E}^{2}-3(\boldsymbol{v}\cdot\boldsymbol{E})^2\big]+\frac{m^{2}}{\epsilon_{\boldsymbol{p}}} - \frac{6\xi-1}{\epsilon_{\boldsymbol{p}}}\left(\frac{a''}{a^{3}}-\frac{a^{\prime 2}}{a^{4}} \right)\\
    &-\frac{6\xi-1}{2\epsilon_{\boldsymbol{p}}^{3}}m^{2}\left(\frac{2a''}{a}-\frac{a^{\prime 2}}{a^{2}}\right)+ \frac{(6\xi-1)^{2}}{2\epsilon_{\boldsymbol{p}}^{3}}\left(\frac{3a^{\prime\prime 2}}{2a^{4}} - \frac{3a''a^{\prime 2}}{a^{5}}-\frac{a^{\mathrm{IV}}}{2a^{3}}+\frac{2a'''a'}{a^{4}} \right)    \bigg\}  \, .\nonumber
\end{align}
Obviously, the first two terms identically vanish since they are odd in momentum, the third term does not give divergent contribution according to Eqs.~\eqref{dimreg-1} and \eqref{dimreg-4}. Thus, the divergent part of the trace does not depend on the electric field. Divergences in the rest of the terms can be extracted applying the dimensional regularization and using Eqs.~\eqref{dimreg-3}--\eqref{dimreg-4}. Then, we obtain the following expression:
\begin{equation}
\label{T_chi_div}
    T_{\chi}^{\mathrm{div}}\!=\!\bigg[\!-\!\frac{m^{4}}{8\pi^{2}}\!-\!(6\xi\!-\!1)\frac{m^{2}}{8\pi^{2}} \frac{a''}{a^{3}}\!+\!\frac{(6\xi\!-\!1)^{2}}{8\pi^{2}}\!\left(\frac{3a^{\prime\prime 2}}{2a^{6}}\! -\! \frac{3a''a^{\prime 2}}{a^{7}}\!-\!\frac{a^{\mathrm{IV}}}{2a^{5}}\!+\!\frac{2a'''a'}{a^{6}} \right)\!      \bigg] \, \frac{1}{\bar{\varepsilon}}\, . 
\end{equation}

It is straightforward to check that this expression is fully canceled by the trace of the effective energy--momentum tensor~\eqref{T_ct} if the coefficients $a_1$, $a_2$, and $a_3$ are given by Eqs.~\eqref{a1}--\eqref{a3}. Thus, we need to add no new counterterms in order to cancel the divergence in the trace of the energy--momentum tensor.

In order to get the finite part of $T_\chi$, we subtract from the integrand in Eq.~\eqref{T_chi_3} the terms in large-$p$ expansion up to $O(\epsilon_{\boldsymbol{p}}^{-3})$ inclusive, shown in Eq.~\eqref{T_chi_exp} and take into account the finite contribution coming from the unity in Eq.~\eqref{dimreg-3}. This leads to the following expression:
\begin{align}
\label{T_chi_reg}
    T_{\chi}^{\mathrm{reg}}&=-\frac{m^{4}}{8\pi^{2}}-(6\xi-1)\frac{m^{2}}{8\pi^{2}}\left(\frac{a''}{a^{3}}-\frac{a^{\prime 2}}{a^{4}} \right) \nonumber\\
    &+ \int \frac{d^{3}\boldsymbol{p}}{(2\pi)^{3}\,a^{4}} \bigg\{ - 4(6\xi-1)\Big[\omega(\eta,\boldsymbol{p})\mathcal{G}(\eta,\boldsymbol{p})+\frac{a'}{a} \mathcal{H}(\eta,\boldsymbol{p}) \Big] \nonumber\\
    &+\frac{2}{\omega(\eta,\boldsymbol{p})} \Big[m^{2}a^2-(6\xi-1) \Big(\frac{a''}{a}-\frac{a^{\prime 2}}{a^{2}} \Big)\Big]\Big[\frac{1}{2}+\mathcal{F}(\eta,\boldsymbol{p})+\mathcal{G}(\eta,\boldsymbol{p})\Big] \nonumber\\
    &+ e\frac{a'}{a}\frac{(\boldsymbol{v}\cdot\boldsymbol{E})}{4\epsilon_{\boldsymbol{p}}^{2}} +\frac{(6\xi-1)e}{2\epsilon_{\boldsymbol{p}}^{2}} \boldsymbol{v}\cdot\Big(\boldsymbol{E}'+2\frac{a'}{a}\boldsymbol{E}\Big)+\frac{(6\xi-1)a^2 e^2}{2\epsilon_{\boldsymbol{p}}^{3}}\big[\boldsymbol{E}^{2}-3(\boldsymbol{v}\cdot\boldsymbol{E})^2\big]\nonumber\\
    &-\frac{m^{2}}{\epsilon_{\boldsymbol{p}}} + \frac{6\xi-1}{\epsilon_{\boldsymbol{p}}}\left(\frac{a''}{a^{3}}-\frac{a^{\prime 2}}{a^{4}} \right) +\frac{6\xi-1}{2\epsilon_{\boldsymbol{p}}^{3}}m^{2}\left(\frac{2a''}{a}-\frac{a^{\prime 2}}{a^{2}}\right)\nonumber\\
    &- \frac{(6\xi-1)^{2}}{2\epsilon_{\boldsymbol{p}}^{3}}\left(\frac{3a^{\prime\prime 2}}{2a^{4}} - \frac{3a''a^{\prime 2}}{a^{5}}-\frac{a^{\mathrm{IV}}}{2a^{3}}+\frac{2a'''a'}{a^{4}} \right)\bigg\}\, .
\end{align}

This expression together with Eq.~\eqref{rho_reg} can be used to derive the pressure of the produced scalar particles, $P_\chi= (\rho_\chi^{\mathrm{reg}}-T_\chi^{\mathrm{reg}})/3$.

\section{Conclusions}
\label{sec:conclusions}

In this work, we describe the production of spinless charged particles (and antiparticles) in a homogeneous electric field in the expanding FLRW universe from the first principles. To this end, we study the particle excitations of a quantum scalar field which evolves on a classical time-dependent background consisting of a scale factor $a(\eta)$ and the electromagnetic vector-potential $\boldsymbol{A}(\eta)$. Because of the absence of static asymptotical states of this system in the infinite past and future we face a typical problem of how to define the vacuum state and particles in the expanding universe. Following the standard procedure~\cite{Birrell-book,Parker-book}, we construct a physically reasonable approximation for the vacuum state---the adiabatic vacuum---in which the modes with very large momenta are almost not excited. However, this procedure is not unique as one can define adiabatic vacua of different adiabatic order and at different moments of time. In particular, in Ref.~\cite{Sobol:2020frh} and in the present work the same physical system was considered while difference is in the choice of adiabatic vacuum which leads to a slightly different results. We argue that the approach presented in this work leads to a system of equations which is more suitable for numerical simulations of the problem.

Throughout the paper, we work with the conformal time in terms of which the FLRW metric is explicitly conformally flat. In a simple case of a conformally coupled massless scalar field in the absence of the electric field, the mode equation admits an exact solution which corresponds to the Bunch--Davies vacuum. Provided that initially the system was in this state, no particle creation occurs at any moment of time in future. This is a manifestation of a conformal symmetry of the physical system under consideration which makes its evolution in FLRW spacetime to be identical to that in the Minkowski spacetime. This symmetry was not respected in the system of equations in Ref.~\cite{Sobol:2020frh}, where definition of the vacuum state is based on the Wentzel--Kramers--Brillouin solution to the mode equation written in physical time.

In a general case, where the conformal symmetry is broken by the particle's mass, nonminimal coupling $\xi\neq 1/6$, and/or the external electric field, the particle creation is described by the system of three quantum Vlasov equations for kinetic functions $\mathcal{F}(\eta,\boldsymbol{p})$, $\mathcal{G}(\eta,\boldsymbol{p})$, and $\mathcal{H}(\eta,\boldsymbol{p})$. The first one is the analog of a classical one-particle distribution function in the Boltzmann kinetic theory while the two others arise as auxiliary dynamical variables which describe the nonlocal in time process of particle creation. 

We study the asymptotical behavior of kinetic functions at large momenta and show that they decrease at large momenta much faster than in the case of Ref.~\cite{Sobol:2020frh}. Indeed, the leading term in the expansion in inverse powers of momentum is $\propto p^{-4}$ for $\mathcal{F}(\eta,\boldsymbol{p})$, $\propto p^{-3}$ for $\mathcal{G}(\eta,\boldsymbol{p})$, and $\propto p^{-2}$ for $\mathcal{H}(\eta,\boldsymbol{p})$ compared to $\propto p^{-2}$, $p^{-2}$, and $p^{-1}$ in the respective cases on Ref.~\cite{Sobol:2020frh}. This makes our new system of equations more suitable for numerical analysis on a lattice in the momentum space.

The powerlike behavior of the kinetic functions at large momenta leads to the UV divergences in the main observables such as the electric current or the energy--momentum tensor of produced particles. We use the dimensional regularization in order to extract the divergent contributions and cancel them by introducing the counterterms to the action. The counterterms coincide with those derived in Ref.~\cite{Sobol:2020frh} and with a well-known results for the scalar quantum electrodynamics obtained by the conventional Green-function approach~\cite{Srednicki-book, Christensen:1976vb, Birrell-book}.
 
The obtained system of quantum Vlasov equations can be further used to describe the dynamics of particle production by a strong electric field in models of inflationary magnetogenesis. Another interesting problem which has a direct physical application within the Standard Model is to describe in a similar quantum kinetic approach the production of charged fermions. The presence of external magnetic field in addition to the electric one makes the dynamics much more interesting and may lead to chirality production via the chiral anomaly~\cite{Adler:1969gk,Bell:1969ts}. The quantum kinetic description of particle production in nonorthogonal electric and magnetic fields is very relevant, e.g., for the axion inflation model; however, it is still missing in the literature.  We plan to address these issues elsewhere.

\vskip.25cm
{\small \noindent \textbf{Acknowledgements} O.\,O.\,S. is grateful to Prof.~Kai Schmitz and all members of Particle Cosmology group for their kind hospitality at the University of M\"{u}nster where the final part of this work was done.}

\section*{Declarations}

{\small \noindent \textbf{Funding} The work was supported by the National Research Foundation of Ukraine (Project No.~2020.02/0062). The work of O.\,O.\,S. was sustained by a Philipp-Schwartz fellowship of the University of M\"{u}nster.}

\vskip.15cm
{\small \noindent \textbf{Conflicts of interest} The authors declare no competing interests.}

\vskip.15cm
{\small \noindent \textbf{Author Contributions} All authors contributed in the writing and research of this paper. A.~L. did computations and wrote the first draft of Sec.~2, 3 and Appendix A.  O.~S. performed computations in Sec.~4 and wrote Sec.~1, 4, and 5. Both authors have checked and approved the manuscript.}

\vskip.15cm
{\small \noindent \textbf{Data availability} No new data were created or analysed in this study.}

\begin{appendices}

\section{Expansion in the inverse powers of momentum}
\label{app-expansion}

In this Appendix, we derive asymptotical expressions for the kinetic functions $\mathcal{F}(\eta,\boldsymbol{p})$, $\mathcal{G}(\eta,\boldsymbol{p})$, and $\mathcal{H}(\eta,\boldsymbol{p})$ in the limit of large momenta. For this, we need the corresponding expansions of the quantities $\omega(\eta,\boldsymbol{p})$ and $Q(\eta,\boldsymbol{p})$ which are the coefficients in the system of quantum Vlasov equations \eqref{eq_F_G_H}:
\begin{equation}
\label{as_o_p}
\omega(\eta,\boldsymbol{p})=p+\omega^{(-1)}+ \mathcal{O}(p^{-3})\, , \qquad \omega^{(-1)}=\frac{m^{2}a^{3}+(6\xi-1)a''}{2ap}\, ,
\end{equation}
\begin{equation}
\label{as_Q_p}
Q(\eta,\boldsymbol{p})=Q^{(-1)}+Q^{(-2)}+ \mathcal{O}(p^{-3})\, ,
\end{equation}
\begin{equation}
\label{Q_p_-1}
Q^{(-1)}=\frac{e\,a^{2}(\boldsymbol{p}\!\cdot\!\boldsymbol{E})}{p^{2}}\, ,	\quad Q^{(-2)}=\frac{1}{p^{2}}\left[m^{2}aa'+\frac{6\xi-1}{2}\left(\frac{a'''}{a}-\frac{a'a''}{a^{2}} \right)    \right] \, .
\end{equation}

Let us represent the total derivative operator with respect to conformal time as the sum of two operators $\hat{L}^{(0)}$ and $\hat{L}^{(-1)}$:
\begin{equation}
\label{d_eta}
\frac{d}{d\eta}=\hat{L}^{(0)}+\hat{L}^{(-1)}\, ,\quad \text{where} \quad \hat{L}^{(0)}=\frac{\partial}{\partial\eta}\, ,\quad \hat{L}^{(-1)}=e\,a^{2}\boldsymbol{E}\frac{\partial}{\partial\boldsymbol{p}}\,.
\end{equation}
Obviously, the operator $\hat{L}^{(0)}$ does not change the asymptotical behavior of a given term at large momenta while $\hat{L}^{(-1)}$ reduces the power of momentum by one.

Further, we represent the kinetic functions as power series in inverse momentum: for $\mathcal{F}(\eta,\boldsymbol{p})$, $\mathcal{G}(\eta,\boldsymbol{p})$, and $\mathcal{H}(\eta,\boldsymbol{p})$ the series starts from the term $\propto p^{-4}$, $\propto p^{-3}$, and $\propto p^{-2}$, respectively. Then, substituting these expansions together with Eqs.~\eqref{as_o_p}--\eqref{d_eta} into the system of quantum Vlasov equations~(\ref{eq_F_G_H}), we obtain the following set of equations:
\begin{align}
\label{eq_1}
    \hat{L}^{(0)}\mathcal{F}^{(-4)}&=Q^{(-1)}\mathcal{G}^{(-3)}\, ,\\
\label{eq_2}
    0&=\frac{1}{2}Q^{(-1)}+2p\,\mathcal{H}^{(-2)}\, ,\\
\label{eq_3}
    0&=\frac{1}{2}Q^{(-2)}+2p\,\mathcal{H}^{(-3)}\, ,\\
\label{eq_4}
    \hat{L}^{(0)}\mathcal{H}^{(-2)}&=-2p\,\mathcal{G}^{(-3)}\, ,\\
\label{eq_5}
    \hat{L}^{(0)}\mathcal{H}^{(-3)}+\hat{L}^{(-1)}\mathcal{H}^{(-2)}&=-2p\,\mathcal{G}^{(-4)}\, .
\end{align}

From Eqs.~\eqref{eq_2} and \eqref{eq_3} we immediately find expressions for the terms $\mathcal{H}^{(-2)}$ and $\mathcal{H}^{(-3)}$:
\begin{align}
\label{H_p_-2}
    \mathcal{H}^{(-2)}&=-\frac{1}{4p}Q^{(-1)}=-\frac{e\,a^{2}(\hat{\boldsymbol{p}}\!\cdot\!\boldsymbol{E})}{4p^{2}}\, ,\\
\label{H_p_-3}
    \mathcal{H}^{(-3)}&=-\frac{1}{4p}Q^{(-2)}=-\frac{1}{4p^{3}}\left[m^{2}aa'+\frac{6\xi-1}{2}\left(\frac{a'''}{a}-\frac{a'a''}{a^{2}} \right)    \right]\, ,
\end{align}
where $\hat{\boldsymbol{p}}=\boldsymbol{p}/p$ is the unit vector in the direction of $\boldsymbol{p}$.

Then, the terms $\mathcal{G}^{(-3)}$ and $\mathcal{G}^{(-4)}$ can be found from Eqs.~\eqref{eq_4} and \eqref{eq_5}, respectively:
\begin{align}
\label{G_p_-3}
    \mathcal{G}^{(-3)}&=-\frac{1}{2p}\hat{L}^{(0)}\mathcal{H}^{(-2)}=\frac{ea^2}{8p^{3}}\,\hat{\boldsymbol{p}}\cdot\Big(\boldsymbol{E}'+2\frac{a'}{a}\boldsymbol{E} \Big) \, ,\\
\label{G_p_-4}
    \mathcal{G}^{(-4)}&=-\frac{1}{2p}\left[\hat{L}^{(0)}\mathcal{H}^{(-3)}+\hat{L}^{(-1)}\mathcal{H}^{(-2)} \right] =\frac{e^{2}a^{4}\big[\boldsymbol{E}^{2}-3(\hat{\boldsymbol{p}}\cdot\boldsymbol{E})^2\big]}{8p^{4}} \nonumber\\
    &+ \frac{1}{8p^{4}}\!\bigg[m^{2}(a^{\prime 2}+aa'') + (6\xi-1)\left(\frac{a^{\mathrm{IV}}}{2a}-\frac{a^{\prime\prime 2}}{2a^{2}}-\frac{a'a'''}{a^{2}} + \frac{a^{\prime 2}a''}{a^{3}} \right)  \bigg]\, .
\end{align}

Finally, the term $\mathcal{F}^{(-4)}$ can be found by solving differential equation \eqref{eq_1}. It is easy to see that the following function is a solution to this equation:
\begin{equation}
\label{F_p_-4}
    \mathcal{F}^{(-4)}=\frac{e^{2}a^{4}(\hat{\boldsymbol{p}}\!\cdot\!\boldsymbol{E})^{2}}{16p^{4}}\, .
\end{equation}

Thus, we derived a few first terms in the Laurent series for the kinetic functions $\mathcal{F}(\eta,\boldsymbol{p})$, $\mathcal{G}(\eta,\boldsymbol{p})$, and $\mathcal{H}(\eta,\boldsymbol{p})$ at large momenta $p\to\infty$. However, it is not convenient to use them in the computations in Sec.~\ref{sec:renormalization} because they lead to spurious infrared divergences in the integrals for physical observables. In order to overcome this problem, it is more convenient to perform expansion in inverse powers of $\epsilon_{\boldsymbol{p}}$ instead of $p$. Therefore, in Eqs.~\eqref{H_p_-2}--\eqref{F_p_-4} we replace $p\to \epsilon_{\boldsymbol{p}}$ in denominators and get Eqs.~\eqref{F-asym}--\eqref{H-asym} in the main text.

\end{appendices}


\end{document}